\documentclass{nature_arxiv}

\usepackage[top=0.8in, bottom=1in, left=0.8in, right=0.8in]{geometry}

\usepackage{graphicx}
\usepackage{amsmath}
\usepackage{amssymb}

\newcommand{\kms}{km\,s$^{-1}$}

\title{Numerous bow shocks in the outer Helix Nebula}

\author{\large Pieter van Dokkum$^{1,2}$,
Roberto Abraham$^{1,3}$,
William P.\ Bowman$^{1}$,
Seery Chen$^{1}$,
Steven R.\ Janssens$^{1}$,
Deborah M.\ Lokhorst$^{1,4}$,
Imad Pasha$^{1}$,
Carter Rhea$^{1}$
\vspace{8pt}}

\begin{document}

\maketitle

\begin{affiliations}
\small
\item Dragonfly Focused Research Organization, 150 Washington Avenue, Suite 201, Santa Fe, NM 87501, USA
\item Department of Astronomy, Yale University, New Haven, CT 06511, USA
\item Department of Astronomy \& Astrophysics, University of Toronto,
50 St.\ George Street, Toronto, ON M5S 3H4, Canada
\item NRC Herzberg Astronomy \& Astrophysics Research Centre, 5071 West Saanich Road, Victoria, BC V9E 2E7, Canada
 
\end{affiliations}


\begin{abstract}

Near the end of their lives, low- and intermediate-mass stars expel metal-enriched material in winds and outflows, ultimately producing planetary nebulae (PNe).\cite{kwok78,iben83} The ejected material is expected to fragment and mix into the interstellar medium (ISM), but this final assimilation step has been difficult to observe directly.\cite{borkowski90,wareing07}
Here we report evidence for this process in the form of twenty-two bow shocks in the eastern outskirts of the Helix Nebula, detected in H$\alpha$ emission with the partially completed MOTHRA telescope.
Unlike the large-scale wind--ISM bow shocks commonly observed around evolved stars,\cite{wareing06,Martin2007,Cox2012} the shocks are compact and associated with individual clumps of gas.
Going outward from the central star, the radius of curvature $R_{\rm c}$
decreases by a factor of $\sim 10^2$ over the radial range $r=0.4$--$1.4$\,pc.
This is accompanied by a morphological transition from thin, well-defined
bows to fuzzy, patchy structures.
We interpret these changes as progressive stripping and fragmentation
of AGB-shell remnants as they interact with the ISM.
The slope of the observed $R_{\rm c}$--$r$ relation implies a loss
of fragment coherence on a timescale of $\approx 10^4$\,yr, providing a rare direct constraint on the time-scale for disruption and entrainment of fragmented stellar ejecta into the ISM.\cite{klein:94,karakas14}

\end{abstract}

The Helix Nebula (NGC\,7293) is one of the closest and brightest PNe, and therefore a benchmark for resolving how the late-stage ejecta of stars couple to their surroundings. Using the Gaia EDR3/DR3 astrometric solution for the central star (WD\,2226$-$210), we adopt a distance of $d = 198.6^{+1.6}_{-1.8}$\,pc.\cite{bailerjones21} Imaging of the Helix has shown it to be highly complex. 
Its bright main nebula comprises an inner disk and a surrounding outer torus, embedded within a larger structure whose upstream side is truncated, consistent with interaction between the expanding AGB ejecta and the ambient interstellar medium (ISM).\cite{odell:04,meaburn:05,zhang:12}
The ionized nebula is threaded by thousands of dense cometary knots and associated molecular material, indicating that much of the ejected material remains in a clumpy, only partially processed phase.\cite{odell:04,hora06}
Deep imaging and spectroscopy have also revealed a bow-shock feature in the faint outer halo, in the direction of
the nebula’s motion through the local ISM.\cite{meaburn:05,meaburn:13} 
Together, these properties make the Helix uniquely suited to place direct constraints on how fragmented stellar ejecta are dispersed and mixed into the ISM —- an essential step in the recycling of mass, dust, and newly synthesized elements in galaxies.\cite{wareing07,karakas14}

The Helix Nebula was observed in the light of
H$\alpha$, [N\,{\sc ii}], and [O\,{\sc iii}] with the partially-built Modular
Optical Telephoto Hyperspectral Robotic Array (MOTHRA).
MOTHRA is an array of high-end telephoto lenses equipped with tiltable
ultra-narrow interference filters, located at the El Sauce Observatory in
Chile. Its design evolved from the Dragonfly Spectral Line Mapper
at New Mexico Skies Observatory.\cite{lokhorst:24,chen:25}
When completed, MOTHRA will have
1140 lenses distributed over 30 mounts, and be
optically-equivalent to a 4.8\,m f/0.08 refractor. 
The data described here are equivalent to $\approx 20$\,min
on-source exposure time with the completed array.

\begin{figure*}[ht]
  \begin{center}
  \includegraphics[width=1.0\linewidth]{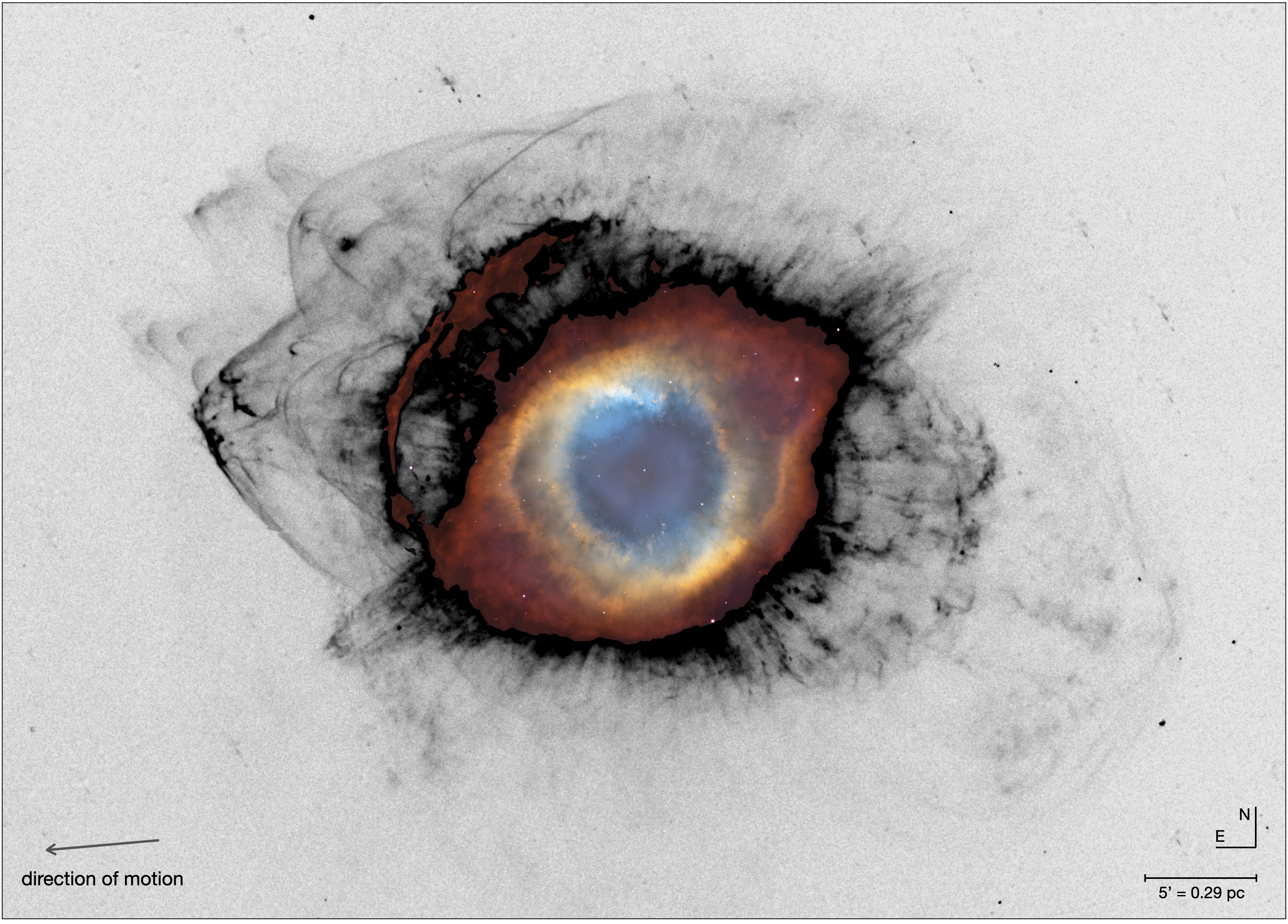}
  \end{center}
    \vspace{-0.3truecm}  
    \caption{\small \textbf{MOTHRA H$\alpha$ imaging of the Helix nebula.} 
The MOTHRA continuum-subtracted H$\alpha$ image is shown with an inverted
grey scale, emphasizing faint outer features. In the bright central
regions a combined Hubble Space Telescope (HST) and Kitt Peak 4m image
is superposed on the MOTHRA
data.\cite{odell:04} The HST image was generated from data in the Advanced Camera
for Surveys F502N ([O\,{\sc iii}]) and F658N (H$\alpha$) filters.
The arrow indicates the Gaia-determined direction of motion of the
central white dwarf with respect to the ambient gas.
The MOTHRA image shows many 
features at large ($\gtrsim 1$\,pc) distances
from the WD that had not been detected in H$\alpha$ before.
The most striking of these
are numerous bow shocks on the east side of the nebula,
where AGB ejecta encounter the ambient ISM at supersonic relative velocities.
The colour Hubble Space Telescope\,/\,Kitt Peak
image is reproduced from NASA, ESA, C.\ R.\ O’Dell (Vanderbilt
University) and M.\ Meixner, P.\ McCullough and G.\ Bacon (Space Telescope
Science Institute).
   }
   \label{helix_ha_large.fig}
    \vspace{-12pt}
\end{figure*}

\begin{figure}[ht]
  \begin{center}
  \includegraphics[width=1.0\linewidth]{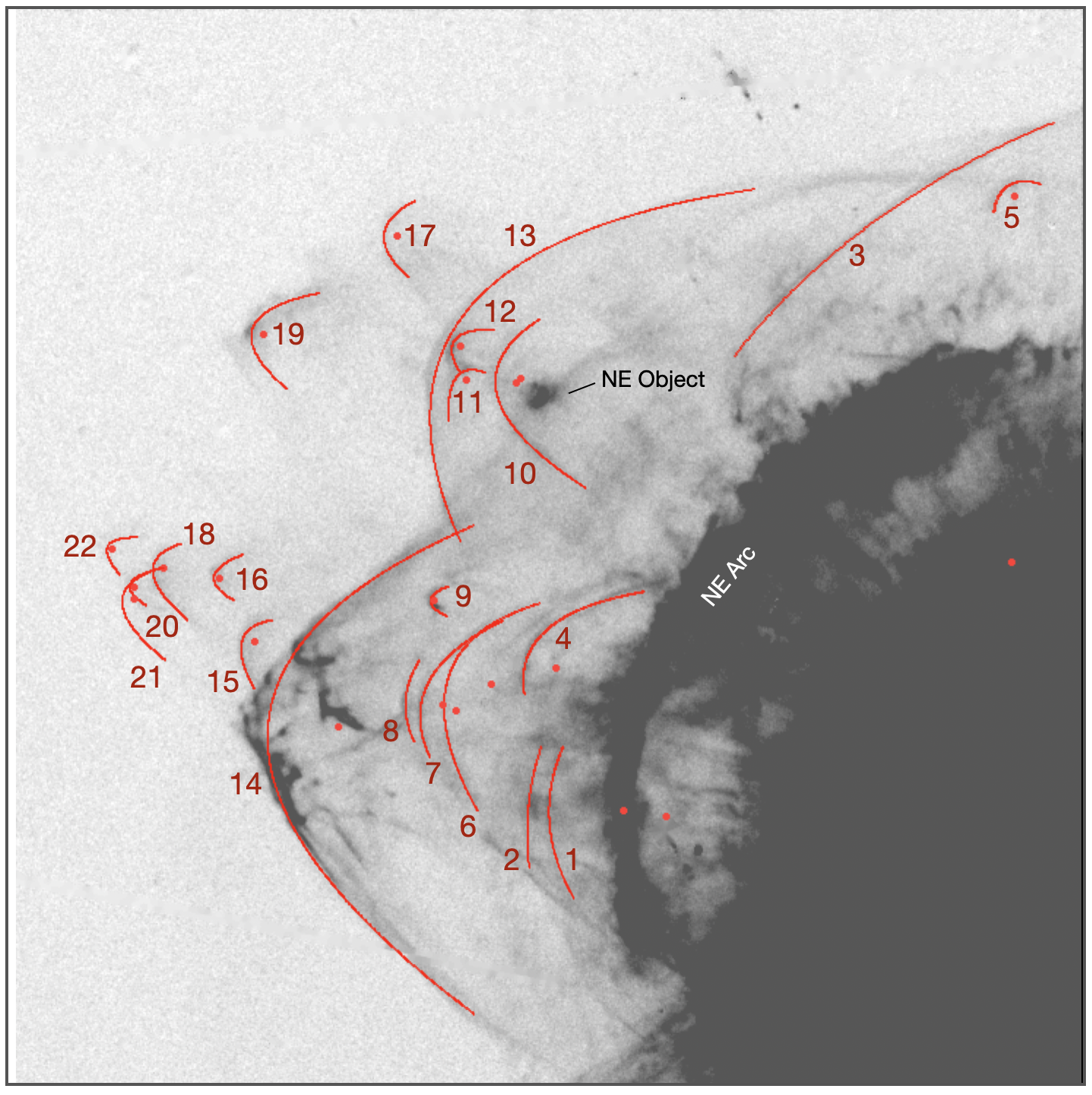}
  \end{center}
    \vspace{-0.3truecm}  
    \caption{\small \textbf{Parabolic profile fits to bow shocks.} 
{\em Left:} Twenty-two complete and partial bow shocks are identified
in the H$\alpha$ image. They are
fit with parabolas, indicated
with the red lines (see Methods). Red dots indicate the foci of the
parabolas; these correspond to the expected approximate locations of the objects
that produce the shocks. The lack of H$\alpha$ detections near the foci indicates
that the objects producing the bows are largely neutral.
The bows are numbered according to the distance
of the apex to the central star. Two known features, the NE Object and the
NE Arc,\cite{odell:04} 
are also marked.
   }
   \label{bow_fit.fig}
    \vspace{-12pt}
\end{figure}

The MOTHRA H$\alpha$
image of the Helix is shown in Fig.\ \ref{helix_ha_large.fig}. It shows
many features that have not been seen in ionized gas before, such as extensions
of the plumes in the northwest and southeast\cite{odell:04,zhang:12} and
turbulent and complex low surface brightness
H$\alpha$ emission in the southwest.\cite{zhang:12}
The most striking feature in the H$\alpha$ image is a forest of arcs and
partial arcs on the eastern side of the nebula.
We identify at least 22 arcs on the eastern side, labeled $1-22$
in Fig.\ \ref{bow_fit.fig} in order of increasing distance from the
central white dwarf.  While the majority of the features appear
to be new discoveries, several can be seen
in previous GALEX and H$\alpha$ images.
Besides the large and
complex feature 14 this includes arcs 3, 10, and 13, among
others.\cite{odell:04,zhang:12,meaburn:13} 
The arcs are undected in [O\,III] and faint
in [N\,II]; from the brightest region of arc 14 we measure
[N\,II]/H$\alpha = 0.06 \pm 0.01$ and [O\,III]/H$\alpha<0.015$ (2$\sigma$).
We also find faint arc-like features on the western side, at a similar distance
from the white dwarf as the much brighter ones in the east.

Following earlier studies we
interpret the eastern features as bow shocks where expanding stellar ejecta 
encounter the ISM at supersonic velocities.\cite{zhang:12,meaburn:13} 
The velocity of the Helix with respect to the ISM is
$\approx36$\,\kms\
toward $95^{\circ}$ East of North in the plane of the sky and $\approx 27$\,\kms\
along the line of sight,\cite{meaburn:13,gaia_dr3}
for a combined ISM velocity of $v_{\rm ISM}\approx 45$\,\kms\ with respect
to the systemic velocity of the nebula. On the eastern side the
shock velocity is the sum of $v_{\rm ISM}$ and the
expansion velocity of the ejecta, $v_{\rm shock,e} \approx |v_{\rm ISM} + v_{\rm exp}|$.
On the western side the ejecta encounter a turbulent
post-shock wake that has passed through, and mixed with,
the PN, $v_{\rm shock,w} \approx |v_{\rm wake} - v_{\rm exp}|$.
The wake velocity is expected to be small with respect to the systemic
velocity of the nebula.
In the case of the well-studied AGB star Mira the processed gas being shed
from the bow shock into the immediate downstream tail lags the star by only
$\sim 10$\,\%;\cite{wareing07,Matthews2008} applying the same
scaling to the Helix gives
$v_{\rm wake} =0-10$\,\kms\ for the flow on the western side.

\begin{figure}[ht]
  \begin{center}
  \includegraphics[width=1.0\linewidth]{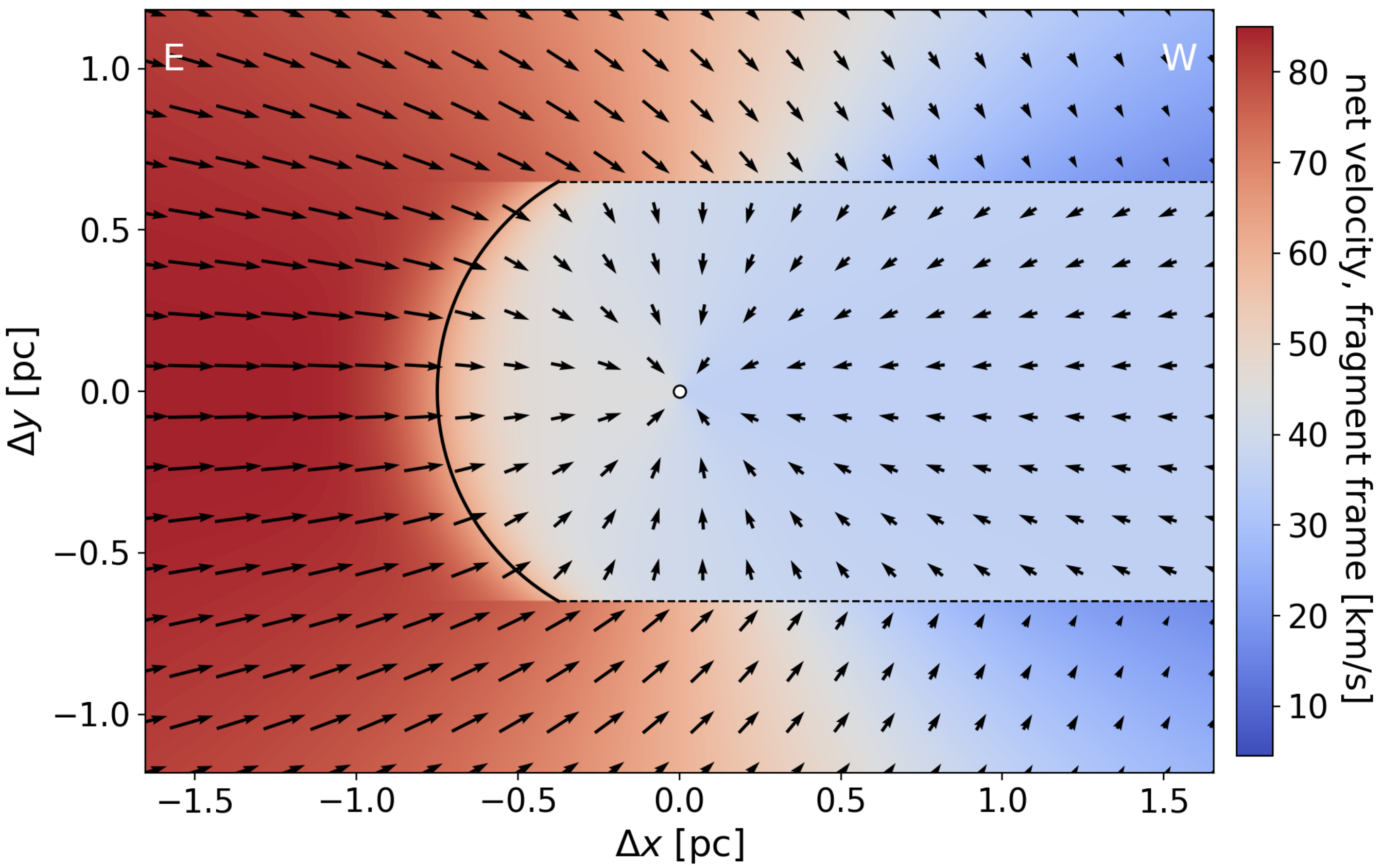}
  \end{center}
    \vspace{-0.5truecm}  
    \caption{\small \textbf{Velocities experienced by fragments.} 
Schematic velocity field around the Helix, for a bulk
velocity with respect to the ISM of $v_{\rm ISM}=45$\,\kms\
eastward, a post-shock flow velocity
of $v_{\rm wake}=5$\,\kms, and a radial expansion velocity
of the ejecta of $v_{\rm exp}=40$\,\kms. 
The highest velocities are found on the east side, where we see the
strong bow shocks.
The scale of the image matches
that of Fig.\ \ref{helix_ha_large.fig}.
   }
   \label{vfield.fig}
    \vspace{-12pt}
\end{figure}

Using the
\textsc{Mappings~V} code\cite{Sutherland2018} we derive
shock velocities on the eastern side
of $v_{\rm shock,e} = 80-90$\,\kms\ from the [O\,III]/H$\alpha$ and
[N\,II]/H$\alpha$ line ratios (see Methods).
Subtracting the $45$\,\kms\ ISM velocity gives an
expansion velocity of the ejecta of $v_{\rm exp}=35-45$\,\kms,
consistent with previously measured H$\alpha$ kinematics in the region
of the brightest bow.\cite{meaburn:13}  
The implied velocity field around the Helix is shown in Fig.\ \ref{vfield.fig}.
Near the E-W axis
the shock velocities are $\sim 80$\,\kms\ in the east and
$\sim 35$\,\kms\ in the west.
For these velocities,
shock models predict H$\alpha$ luminosities that are
1--2 orders of magnitude fainter in the west than in the east,
consistent with the appearance
of the bows on the two sides of the Helix (see Methods).

The expansion velocity of $v_{\rm exp}=35$--$45$\,\kms\ implies a dynamical age of
the clumps of 20,000\,--\,30,000\,yr at $r\sim 1$\,pc, predating the formation of the
planetary nebula $\sim 12{,}000$\,yr ago.\cite{meaburn:08}
The material therefore most likely belongs to an older circumstellar envelope that
was ejected during the late AGB phase, and has now fragmented into many individual clumps.
This interpretation is strengthened by the fact that the bows lie at approximately the same
radius as the $\sim 40'$ outer \textit{WISE} 12$\mu$m halo, which has been associated with dust from an AGB wind.\cite{zhang:12}
Fast winds and outflows, with velocities that can exceed the canonical $\sim 5-20$\,\kms\ range
of the main AGB phase,\cite{hofnerolofsson:18}
are commonly observed during the late-AGB and early post-AGB
phases.\cite{balickfrank:02,sahai:01} These flows are often bipolar rather than
isotropic,\cite{bujarrabal:01}
which may explain why there appears to be a preferred
axis connecting the strong bows in the east / northeast
of the Helix to the weak bows in the west / southwest.


We fit the bow morphologies with a range of functional forms:
parabolas, hyperbolas, ellipses, and Wilkinoids.\cite{Wilkin1996}
We adopt the parabolic fits
as our fiducial model, as they provide a reasonable
description of the data and retain the same functional form under projection.\cite{tarango:18}
Since many of the structures do not show a complete bow, and
some are broader and less sharply bounded than ideal
thin-shell bow shocks, we do not attach direct dynamical significance to the adopted
fit family itself. Instead, we use the fits to extract geometric quantities
(in particular the characteristic curvature scale), and use the
variation between fit families as an estimate of the systematic uncertainty.
The fitting procedure is detailed in the Methods section, with
the results shown in Fig.\ \ref{bow_fit.fig}. 
Most of the 22 bows are reasonably well fit by a parabola, although the wings
are often closer to hyperbolic.\cite{tarango:18}
The brightest bow (14) is 
more sharply peaked than the model curve;
inspection of the region near the apex shows that it is broken up in a complex
network of smaller shocks.

The foci of the parabolic fits are indicated with red dots
in Fig.\ \ref{bow_fit.fig}. These are the approximate locations of the shell
fragments that produce the shocks,
although the exact location depends on the 3D orientation and shape
of the bows.\cite{tarango:18}
There is generally nothing visible in H$\alpha$, [O\,III], or [N\,II] at or near
these locations.
The lack of detected emission counterparts at most of
the inferred obstacle locations is consistent with the fragments being largely neutral. 
Our observations thus represent a new observational window on the mixing of AGB\,/\,PN ejecta into the ISM, in which the fragments are identified not by their intrinsic emission but by the shocks that they drive.

\begin{figure*}[htb]
  \begin{center}
  \includegraphics[width=0.8\linewidth]{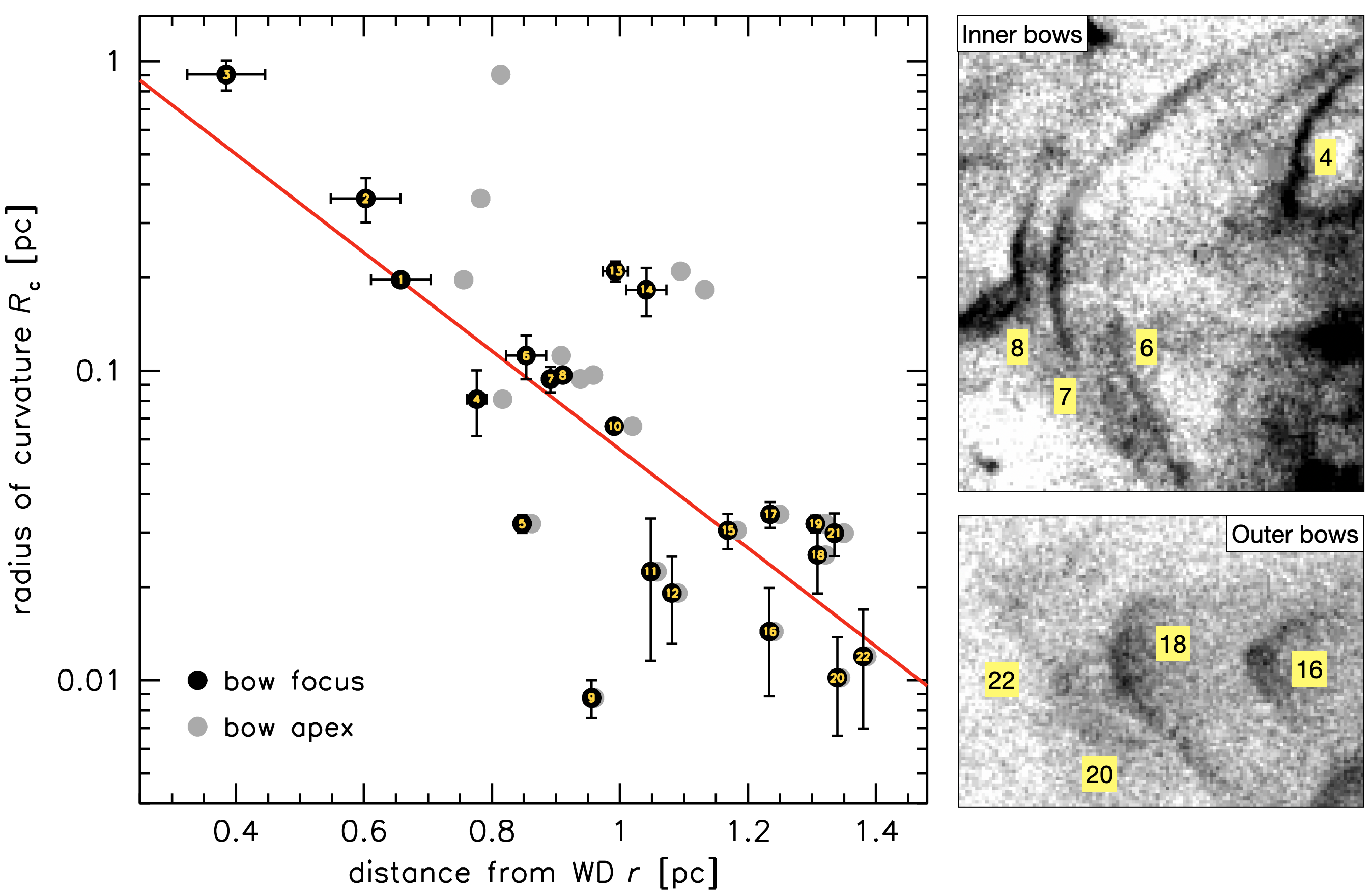}
  \end{center}
    \vspace{-0.3truecm}  
    \caption{\small \textbf{Relation between the size and morphology of bow shocks and their distance from the central star.} 
    {\em Main panel:}
    Radius of curvature $R_{\rm c}$, determined from fitting parabolas to the bows,
    versus distance from the central white dwarf $r$. Grey points indicate the location of the apex of the shock and black points indicate the focus, that is,
    the approximate location of the object that is causing the shock. Yellow numbers identify the bows, ordered by the distance of the apex from the white dwarf. There is a strong dependence, with shocks in the outer parts being smaller.
    The red line is a fit of the form $\log R_{\rm c} = 0.34 - 1.59 r$, corresponding
    to an e-folding length of 0.27\,pc. {\em Right panels:} Representative morphologies of shocks at a distance $r\approx 0.8$\,pc from the white dwarf (top) and at $r\approx 1.3$\,pc (bottom). As $r$ increases, there is an evolution from large, thin, well-defined  shocks (4, 6, 7, 8) to small, broad features (16, 18, 20, 22) that we interpret as a sequence of shell fragment disruption.
   }
   \label{R0r.fig}
    \vspace{-2pt}
\end{figure*}

The nature of the bows changes systematically with distance from the white dwarf:
close to the star they are large, thin, and well-defined, whereas in the outskirts they
are smaller and fuzzier. To quantify this trend without assuming a particular
steady-state bow-shock solution, we characterize each structure by the radius of
curvature at its apex $R_{\rm c}$, as measured from the best-fitting parabola.
The relation between $R_{\rm c}$ and the distance from the white dwarf,
$r$, is shown in Fig.\ \ref{R0r.fig}.
The characteristic size of the bows decreases by two orders of magnitude
over the radial range $0.4\,{\rm pc} \lesssim r \lesssim 1.4\,{\rm pc}$.
A log-linear fit gives $\log R_{\rm c} = 0.34 - 1.59\,r$, corresponding to an
e-folding length of $0.27$\,pc.

Because $R_{\rm c}$ is a purely geometric quantity, it does not by itself specify
the detailed momentum balance within the flow.
It does, however, show that the spatial scale of the coherent bow-forming obstacle
decreases strongly with outward distance.
This geometric trend is accompanied by a systematic morphological transition:
the inner bows are thin and sharply bounded, while the outer structures are broader,
more irregular, and increasingly clumpy.
Taken together, these changes suggest progressive stripping and fragmentation of
the dense AGB-shell remnants as they interact with the ambient medium.
As material is ablated from the fragments and mixed into the surrounding flow,
the surviving dense heads become smaller and more porous, and a larger fraction of
the H$\alpha$ emission likely arises in fragment-associated, mass-loaded mixed gas
rather than in a geometrically thin, well-defined forward shock.\cite{Hartquist1986,Pittard2001,klein:94}

In this interpretation, the bows are both signposts of the fragments and agents
of their destruction.
The shocks are powered by the relative kinetic energy of the fragments and the ambient flow,
and their presence implies ongoing momentum transfer, ablation, and mixing.\cite{Hartquist1986,klein:94}
Interpreting the radial locations of the bows, $r_i$, as a time sequence for outward
evolution, $t_i \approx r_i/v_{\rm exp}$,
the slope of the $R_{\rm c}$--$r$ relation implies a characteristic time scale for the
loss of coherent bow structure.
For the expansion velocity derived from the shock velocities in the east,
$v_{\rm exp} \sim 40$\,\kms, the curvature scale declines with an e-folding time
$\tau_{R_{\rm c}} \sim 7\times 10^3$\,yr.
We therefore infer that the bow-forming AGB shell fragments are disrupted on a
time scale of order $10^4$\,yr.
This estimate should be interpreted as the survival time of the coherent dense
fragment/bow system, rather than as a direct measurement of a specific momentum
or mass-loss rate.

Classical AGB--ISM bow shocks such as Mira trace a single, wind-driven stand-off interaction centered on the mass-losing star,\cite{Martin2007}
and far-infrared surveys show that such global wind--ISM interaction structures are common around evolved stars.\cite{Cox2012} 
However, these long-lived\cite{Martin2007} ($\sim 10^5$\,yr)
large-scale bows primarily map where the wind meets the ISM; they do not directly constrain how fragmented ejecta are ultimately assimilated.
In the Helix outer halo we instead resolve numerous compact bow shocks with no luminous source at their foci, implying that the AGB--ISM interaction has
fragmented into dense, line-dark obstacles that are progressively ablated and entrained. The $\sim 10^4$\,yr characteristic disruption time for the fragments
can therefore be interpreted as the relevant time-scale for recycling of late AGB
ejecta into the ISM.



More broadly, stellar mass loss is a major channel by which galaxies recycle gas, metals,
and dust back into the ISM,\cite{LeitnerKravtsov2011}
yet the efficiency and duration of the final, fragment-driven AGB\,--\,ISM assimilation step
remain poorly constrained observationally.\cite{Martin2007,Matthews2013AJ}
Galaxy formation simulations therefore rely on subgrid turbulent mixing and diffusion prescriptions to represent unresolved transport.\cite{Wadsley2008,Shen2010,Rennehan2019}
Our empirically inferred $\sim10^4$\,yr disruption time implies that once AGB ejecta are fragmented and exposed to the diffuse medium they lose their coherent
identity rapidly, providing a benchmark for models of recycling and feedback.

Our conclusions can be tested and extended in several ways.
The  Helix is a fairly typical PN,
and we should see similar fragment-driven bow shocks
in the outskirts of other PNe. Because of the steep relation between H$\alpha$
luminosity and shock velocity such features will be most readily
detected when the bulk motion of the nebula with respect
to the ISM exceeds $\sim 40$\,\kms. When more examples are found,
it will be interesting to see if the
mixing timescale depends on the shock velocities.
Such observations will be within easy reach of the completed MOTHRA. Furthermore,
by analogy with the molecular cometary knots in the
Helix,\cite{Huggins1992,Huggins2002,Matsuura2009,Hora2006}
the H$\alpha$-dark clumps whose presence is inferred from
the shocks
could be detectable in CO rotational lines and, particularly,
in the H$_2$ 1--0 S(1) 2.12\,$\mu$m line.

\vspace{1cm}

\bibliography{master_0318}

\begin{thebibliography}{10}
\expandafter\ifx\csname url\endcsname\relax
  \def\url#1{\texttt{#1}}\fi
\expandafter\ifx\csname urlprefix\endcsname\relax\def\urlprefix{URL }\fi
\providecommand{\bibinfo}[2]{#2}
\providecommand{\eprint}[2][]{\url{#2}}

\bibitem{kwok78}
\bibinfo{author}{Kwok, S.}, \bibinfo{author}{Purton, C.~R.} \&
  \bibinfo{author}{FitzGerald, P.~M.}
\newblock \bibinfo{title}{On the origin of planetary nebulae}.
\newblock \emph{\bibinfo{journal}{The Astrophysical Journal Letters}}
  \textbf{\bibinfo{volume}{219}}, \bibinfo{pages}{L125--L127}
  (\bibinfo{year}{1978}).

\bibitem{iben83}
\bibinfo{author}{Iben, I.} \& \bibinfo{author}{Renzini, A.}
\newblock \bibinfo{title}{Asymptotic giant branch evolution and beyond}.
\newblock \emph{\bibinfo{journal}{Annual Review of Astronomy and Astrophysics}}
  \textbf{\bibinfo{volume}{21}}, \bibinfo{pages}{271--342}
  (\bibinfo{year}{1983}).

\bibitem{borkowski90}
\bibinfo{author}{Borkowski, K.~J.}, \bibinfo{author}{Sarazin, C.~L.} \&
  \bibinfo{author}{Soker, N.}
\newblock \bibinfo{title}{Interaction of planetary nebulae with the
  interstellar medium}.
\newblock \emph{\bibinfo{journal}{The Astrophysical Journal}}
  \textbf{\bibinfo{volume}{360}}, \bibinfo{pages}{173--183}
  (\bibinfo{year}{1990}).

\bibitem{wareing07}
\bibinfo{author}{Wareing, C.~J.}, \bibinfo{author}{Zijlstra, A.~A.} \&
  \bibinfo{author}{O'Brien, T.~J.}
\newblock \bibinfo{title}{Vortices in the wakes of asymptotic giant branch
  stars}.
\newblock \emph{\bibinfo{journal}{The Astrophysical Journal Letters}}
  \textbf{\bibinfo{volume}{660}}, \bibinfo{pages}{L129--L132}
  (\bibinfo{year}{2007}).

\bibitem{wareing06}
\bibinfo{author}{Wareing, C.~J.} \emph{et~al.}
\newblock \bibinfo{title}{Detached shells as tracers of asymptotic giant
  branch--interstellar medium bow shocks}.
\newblock \emph{\bibinfo{journal}{Monthly Notices of the Royal Astronomical
  Society: Letters}} \textbf{\bibinfo{volume}{372}}, \bibinfo{pages}{L63--L67}
  (\bibinfo{year}{2006}).

\bibitem{Martin2007}
\bibinfo{author}{Martin, D.~C.} \emph{et~al.}
\newblock \bibinfo{title}{A turbulent wake as a tracer of 30,000 years of
  mira's mass loss history}.
\newblock \emph{\bibinfo{journal}{Nature}} \textbf{\bibinfo{volume}{448}},
  \bibinfo{pages}{780--783} (\bibinfo{year}{2007}).

\bibitem{Cox2012}
\bibinfo{author}{Cox, N. L.~J.} \emph{et~al.}
\newblock \bibinfo{title}{A far-infrared survey of bow shocks and detached
  shells around agb stars and red supergiants}.
\newblock \emph{\bibinfo{journal}{Astronomy \& Astrophysics}}
  \textbf{\bibinfo{volume}{537}}, \bibinfo{pages}{A35} (\bibinfo{year}{2012}).
\newblock \eprint{1110.5486}.

\bibitem{klein:94}
\bibinfo{author}{Klein, R.~I.}, \bibinfo{author}{McKee, C.~F.} \&
  \bibinfo{author}{Colella, P.}
\newblock \bibinfo{title}{On the hydrodynamic interaction of shock waves with
  interstellar clouds. 1: Nonradiative shocks in small clouds}.
\newblock \emph{\bibinfo{journal}{Astrophysical Journal}}
  \textbf{\bibinfo{volume}{420}}, \bibinfo{pages}{213--236}
  (\bibinfo{year}{1994}).

\bibitem{karakas14}
\bibinfo{author}{Karakas, A.~I.} \& \bibinfo{author}{Lattanzio, J.~C.}
\newblock \bibinfo{title}{The dawes review 2: Nucleosynthesis and stellar
  yields of low- and intermediate-mass single stars}.
\newblock \emph{\bibinfo{journal}{Publications of the Astronomical Society of
  Australia}} \textbf{\bibinfo{volume}{31}}, \bibinfo{pages}{e030}
  (\bibinfo{year}{2014}).

\bibitem{bailerjones21}
\bibinfo{author}{Bailer-Jones, C. A.~L.}, \bibinfo{author}{Rybizki, J.},
  \bibinfo{author}{Fouesneau, M.}, \bibinfo{author}{Demleitner, M.} \&
  \bibinfo{author}{Andrae, R.}
\newblock \bibinfo{title}{Estimating distances from parallaxes. v. geometric
  and photogeometric distances to 1.47 billion stars in gaia early data release
  3}.
\newblock \emph{\bibinfo{journal}{The Astronomical Journal}}
  \textbf{\bibinfo{volume}{161}}, \bibinfo{pages}{147} (\bibinfo{year}{2021}).

\bibitem{odell:04}
\bibinfo{author}{{O'Dell}, C.~R.}, \bibinfo{author}{{McCullough}, P.~R.} \&
  \bibinfo{author}{{Meixner}, M.}
\newblock \bibinfo{title}{{Unraveling the Helix Nebula: Its Structure and
  Knots}}.
\newblock \emph{\bibinfo{journal}{\aj}} \textbf{\bibinfo{volume}{128}},
  \bibinfo{pages}{2339--2356} (\bibinfo{year}{2004}).
\newblock \eprint{astro-ph/0407556}.

\bibitem{meaburn:05}
\bibinfo{author}{Meaburn, J.} \emph{et~al.}
\newblock \bibinfo{title}{The creation of the helix planetary nebula (ngc 7293)
  by multiple events}.
\newblock \emph{\bibinfo{journal}{Monthly Notices of the Royal Astronomical
  Society}} \textbf{\bibinfo{volume}{360}}, \bibinfo{pages}{963--976}
  (\bibinfo{year}{2005}).

\bibitem{zhang:12}
\bibinfo{author}{{Zhang}, Y.}, \bibinfo{author}{{Hsia}, C.-H.} \&
  \bibinfo{author}{{Kwok}, S.}
\newblock \bibinfo{title}{{Discovery of a Halo around the Helix Nebula NGC 7293
  in the WISE All-sky Survey}}.
\newblock \emph{\bibinfo{journal}{\apj}} \textbf{\bibinfo{volume}{755}},
  \bibinfo{pages}{53} (\bibinfo{year}{2012}).
\newblock \eprint{1207.4606}.

\bibitem{hora06}
\bibinfo{author}{Hora, J.~L.}, \bibinfo{author}{Latter, W.~B.},
  \bibinfo{author}{Smith, H.~A.} \& \bibinfo{author}{Marengo, M.}
\newblock \bibinfo{title}{Infrared observations of the helix planetary nebula}.
\newblock \emph{\bibinfo{journal}{The Astrophysical Journal}}
  \textbf{\bibinfo{volume}{652}}, \bibinfo{pages}{426--441}
  (\bibinfo{year}{2006}).

\bibitem{meaburn:13}
\bibinfo{author}{{Meaburn}, J.}, \bibinfo{author}{{Boumis}, P.} \&
  \bibinfo{author}{{Akras}, S.}
\newblock \bibinfo{title}{{The bow-shock and high-speed jet in the faint, 40
  arcmin diameter, outer halo of the evolved Helix planetary nebula (NGC
  7293)}}.
\newblock \emph{\bibinfo{journal}{\mnras}} \textbf{\bibinfo{volume}{435}},
  \bibinfo{pages}{3462--3468} (\bibinfo{year}{2013}).
\newblock \eprint{1308.5460}.

\bibitem{lokhorst:24}
\bibinfo{author}{{Lokhorst}, D.~M.} \emph{et~al.}
\newblock \bibinfo{title}{{Realizing the potential of the Dragonfly Spectral
  Line Mapper: calibration methods and on-sky performance}}.
\newblock In \bibinfo{editor}{{Marshall}, H.~K.},
  \bibinfo{editor}{{Spyromilio}, J.} \& \bibinfo{editor}{{Usuda}, T.} (eds.)
  \emph{\bibinfo{booktitle}{Ground-based and Airborne Telescopes X}}, vol.
  \bibinfo{volume}{13094} of \emph{\bibinfo{series}{Society of Photo-Optical
  Instrumentation Engineers (SPIE) Conference Series}},
  \bibinfo{pages}{130942N} (\bibinfo{year}{2024}).

\bibitem{chen:25}
\bibinfo{author}{{Chen}, S.} \emph{et~al.}
\newblock \bibinfo{title}{{First Light with the 120-lens Dragonfly Spectral
  Line Mapper}}.
\newblock \emph{\bibinfo{journal}{\pasp}} \textbf{\bibinfo{volume}{137}},
  \bibinfo{pages}{084103} (\bibinfo{year}{2025}).

\bibitem{gaia_dr3}
\bibinfo{author}{{Gaia Collaboration}}, \bibinfo{author}{{Brown}, A. G.~A.}
  \emph{et~al.}
\newblock \bibinfo{title}{Gaia early data release 3: Summary of the contents
  and survey properties}.
\newblock \emph{\bibinfo{journal}{Astronomy \& Astrophysics}}
  \textbf{\bibinfo{volume}{649}}, \bibinfo{pages}{A1} (\bibinfo{year}{2021}).

\bibitem{Matthews2008}
\bibinfo{author}{Matthews, L.~D.}, \bibinfo{author}{Libert, Y.},
  \bibinfo{author}{G{\'e}rard, E.}, \bibinfo{author}{Le~Bertre, T.} \&
  \bibinfo{author}{Reid, M.~J.}
\newblock \bibinfo{title}{The discovery of an h i counterpart to mira's
  far-ultraviolet-emitting tail}.
\newblock \emph{\bibinfo{journal}{The Astrophysical Journal}}
  \textbf{\bibinfo{volume}{684}}, \bibinfo{pages}{603--615}
  (\bibinfo{year}{2008}).

\bibitem{Sutherland2018}
\bibinfo{author}{Sutherland, R.~S.} \& \bibinfo{author}{Dopita, M.~A.}
\newblock \bibinfo{title}{Mappings~v: A new shock and photoionization code}.
\newblock \emph{\bibinfo{journal}{Astrophys. J. Suppl.}}
  \textbf{\bibinfo{volume}{237}}, \bibinfo{pages}{12} (\bibinfo{year}{2018}).

\bibitem{meaburn:08}
\bibinfo{author}{Meaburn, J.}, \bibinfo{author}{L{\'o}pez, J.~A.} \&
  \bibinfo{author}{Richer, M.~G.}
\newblock \bibinfo{title}{Optical line profiles of the helix planetary nebula
  (ngc 7293) to large radii}.
\newblock \emph{\bibinfo{journal}{Monthly Notices of the Royal Astronomical
  Society}} \textbf{\bibinfo{volume}{384}}, \bibinfo{pages}{497--503}
  (\bibinfo{year}{2008}).

\bibitem{hofnerolofsson:18}
\bibinfo{author}{H{\"o}fner, S.} \& \bibinfo{author}{Olofsson, H.}
\newblock \bibinfo{title}{Mass loss of stars on the asymptotic giant branch.
  mechanisms, models and measurements}.
\newblock \emph{\bibinfo{journal}{The Astronomy and Astrophysics Review}}
  \textbf{\bibinfo{volume}{26}}, \bibinfo{pages}{1} (\bibinfo{year}{2018}).

\bibitem{balickfrank:02}
\bibinfo{author}{{Balick}, B.} \& \bibinfo{author}{{Frank}, A.}
\newblock \bibinfo{title}{{Shapes and Shaping of Planetary Nebulae}}.
\newblock \emph{\bibinfo{journal}{\araa}} \textbf{\bibinfo{volume}{40}},
  \bibinfo{pages}{439--486} (\bibinfo{year}{2002}).

\bibitem{sahai:01}
\bibinfo{author}{Sahai, R.}
\newblock \bibinfo{title}{Hst imaging of proto-planetary nebulae and very young
  planetary nebulae---towards a new understanding of their formation}.
\newblock In \emph{\bibinfo{booktitle}{Post-AGB Objects as a Phase of Stellar
  Evolution}}, vol. \bibinfo{volume}{265} of
  \emph{\bibinfo{series}{Astrophysics and Space Science Library}},
  \bibinfo{pages}{53--63} (\bibinfo{publisher}{Springer},
  \bibinfo{year}{2001}).

\bibitem{bujarrabal:01}
\bibinfo{author}{Bujarrabal, V.}, \bibinfo{author}{Castro-Carrizo, A.},
  \bibinfo{author}{Alcolea, J.} \& \bibinfo{author}{S{\'a}nchez~Contreras, C.}
\newblock \bibinfo{title}{Mass, linear momentum and kinetic energy of bipolar
  flows in protoplanetary nebulae}.
\newblock \emph{\bibinfo{journal}{Astronomy \& Astrophysics}}
  \textbf{\bibinfo{volume}{377}}, \bibinfo{pages}{868--897}
  (\bibinfo{year}{2001}).

\bibitem{Wilkin1996}
\bibinfo{author}{Wilkin, F.~P.}
\newblock \bibinfo{title}{Exact analytic solutions for stellar wind bow
  shocks}.
\newblock \emph{\bibinfo{journal}{Astrophys. J. Lett.}}
  \textbf{\bibinfo{volume}{459}}, \bibinfo{pages}{L31--L34}
  (\bibinfo{year}{1996}).

\bibitem{tarango:18}
\bibinfo{author}{{Tarango-Yong}, J.~A.} \& \bibinfo{author}{{Henney}, W.~J.}
\newblock \bibinfo{title}{{True versus apparent shapes of bow shocks}}.
\newblock \emph{\bibinfo{journal}{\mnras}} \textbf{\bibinfo{volume}{477}},
  \bibinfo{pages}{2431--2454} (\bibinfo{year}{2018}).
\newblock \eprint{1712.02300}.

\bibitem{Hartquist1986}
\bibinfo{author}{Hartquist, T.~W.}, \bibinfo{author}{Dyson, J.~E.},
  \bibinfo{author}{Pettini, M.} \& \bibinfo{author}{Smith, L.~J.}
\newblock \bibinfo{title}{Mass-loaded astronomical flows -- i. general
  principles and their application to rcw 58}.
\newblock \emph{\bibinfo{journal}{Monthly Notices of the Royal Astronomical
  Society}} \textbf{\bibinfo{volume}{221}}, \bibinfo{pages}{715--726}
  (\bibinfo{year}{1986}).

\bibitem{Pittard2001}
\bibinfo{author}{Pittard, J.~M.}, \bibinfo{author}{Hartquist, T.~W.} \&
  \bibinfo{author}{Dyson, J.~E.}
\newblock \bibinfo{title}{Self-similar evolution of wind-blown bubbles with
  mass loading by hydrodynamic ablation}.
\newblock \emph{\bibinfo{journal}{Astronomy \& Astrophysics}}
  \textbf{\bibinfo{volume}{373}}, \bibinfo{pages}{1043--1055}
  (\bibinfo{year}{2001}).

\bibitem{LeitnerKravtsov2011}
\bibinfo{author}{Leitner, S.~N.} \& \bibinfo{author}{Kravtsov, A.~V.}
\newblock \bibinfo{title}{Fuel efficient galaxies: Sustaining star formation
  with stellar mass loss}.
\newblock \emph{\bibinfo{journal}{The Astrophysical Journal}}
  \textbf{\bibinfo{volume}{734}}, \bibinfo{pages}{48} (\bibinfo{year}{2011}).
\newblock \eprint{1011.1252}.

\bibitem{Matthews2013AJ}
\bibinfo{author}{Matthews, L.~D.}, \bibinfo{author}{Le~Bertre, T.},
  \bibinfo{author}{G{\'e}rard, E.} \& \bibinfo{author}{Johnson, M.~C.}
\newblock \bibinfo{title}{An h i imaging survey of asymptotic giant branch
  stars}.
\newblock \emph{\bibinfo{journal}{The Astronomical Journal}}
  \textbf{\bibinfo{volume}{145}}, \bibinfo{pages}{97} (\bibinfo{year}{2013}).
\newblock \eprint{1301.7429}.

\bibitem{Wadsley2008}
\bibinfo{author}{Wadsley, J.~W.}, \bibinfo{author}{Veeravalli, G.} \&
  \bibinfo{author}{Couchman, H. M.~P.}
\newblock \bibinfo{title}{On the treatment of entropy mixing in numerical
  cosmology}.
\newblock \emph{\bibinfo{journal}{Monthly Notices of the Royal Astronomical
  Society}} \textbf{\bibinfo{volume}{387}}, \bibinfo{pages}{427--438}
  (\bibinfo{year}{2008}).

\bibitem{Shen2010}
\bibinfo{author}{Shen, S.}, \bibinfo{author}{Wadsley, J.} \&
  \bibinfo{author}{Stinson, G.}
\newblock \bibinfo{title}{The enrichment of the intergalactic medium with
  adiabatic feedback -- i. metal cooling and metal diffusion}.
\newblock \emph{\bibinfo{journal}{Monthly Notices of the Royal Astronomical
  Society}} \textbf{\bibinfo{volume}{407}}, \bibinfo{pages}{1581--1596}
  (\bibinfo{year}{2010}).
\newblock \eprint{0910.5956}.

\bibitem{Rennehan2019}
\bibinfo{author}{Rennehan, D.}, \bibinfo{author}{Babul, A.},
  \bibinfo{author}{Hopkins, P.~F.}, \bibinfo{author}{Dav{\'e}, R.} \&
  \bibinfo{author}{Moa, B.}
\newblock \bibinfo{title}{Dynamic localized turbulent diffusion and its impact
  on the galactic ecosystem}.
\newblock \emph{\bibinfo{journal}{Monthly Notices of the Royal Astronomical
  Society}} \textbf{\bibinfo{volume}{483}}, \bibinfo{pages}{3810--3831}
  (\bibinfo{year}{2019}).

\bibitem{Huggins1992}
\bibinfo{author}{Huggins, P.~J.}, \bibinfo{author}{Bachiller, R.},
  \bibinfo{author}{Cox, P.} \& \bibinfo{author}{Forveille, T.}
\newblock \bibinfo{title}{Co in the cometary globules of the helix nebula}.
\newblock \emph{\bibinfo{journal}{The Astrophysical Journal Letters}}
  \textbf{\bibinfo{volume}{401}}, \bibinfo{pages}{L43--L46}
  (\bibinfo{year}{1992}).

\bibitem{Huggins2002}
\bibinfo{author}{Huggins, P.~J.} \emph{et~al.}
\newblock \bibinfo{title}{High-resolution co and h$_2$ molecular line imaging
  of a cometary globule in the helix nebula}.
\newblock \emph{\bibinfo{journal}{The Astrophysical Journal Letters}}
  \textbf{\bibinfo{volume}{573}}, \bibinfo{pages}{L55--L58}
  (\bibinfo{year}{2002}).
\newblock \eprint{astro-ph/0205516}.

\bibitem{Matsuura2009}
\bibinfo{author}{Matsuura, M.} \emph{et~al.}
\newblock \bibinfo{title}{A ``firework'' of h$_2$ knots in the planetary nebula
  ngc 7293 (the helix nebula)}.
\newblock \emph{\bibinfo{journal}{The Astrophysical Journal}}
  \textbf{\bibinfo{volume}{700}}, \bibinfo{pages}{1067--1077}
  (\bibinfo{year}{2009}).
\newblock \eprint{0906.2870}.

\bibitem{Hora2006}
\bibinfo{author}{Hora, J.~L.}, \bibinfo{author}{Latter, W.~B.},
  \bibinfo{author}{Smith, H.~A.} \& \bibinfo{author}{Marengo, M.}
\newblock \bibinfo{title}{Infrared observations of the helix planetary nebula}.
\newblock \emph{\bibinfo{journal}{The Astrophysical Journal}}
  \textbf{\bibinfo{volume}{652}}, \bibinfo{pages}{426--441}
  (\bibinfo{year}{2006}).
\newblock \eprint{astro-ph/0607541}.

\bibitem{balick87}
\bibinfo{author}{Balick, B.}
\newblock \bibinfo{title}{Evolution of planetary nebulae. i. structures,
  ionizations, and morphological sequences}.
\newblock \emph{\bibinfo{journal}{The Astronomical Journal}}
  \textbf{\bibinfo{volume}{94}}, \bibinfo{pages}{671--678}
  (\bibinfo{year}{1987}).

\bibitem{corradi96}
\bibinfo{author}{Corradi, R. L.~M.}, \bibinfo{author}{Manso, R.},
  \bibinfo{author}{Mampaso, A.} \& \bibinfo{author}{Schwarz, H.~E.}
\newblock \bibinfo{title}{Unveiling low-ionization microstructures in planetary
  nebulae}.
\newblock \emph{\bibinfo{journal}{Astronomy \& Astrophysics}}
  \textbf{\bibinfo{volume}{313}}, \bibinfo{pages}{913--923}
  (\bibinfo{year}{1996}).

\bibitem{goncalves01}
\bibinfo{author}{Gon{\c{c}}alves, D.~R.}, \bibinfo{author}{Corradi, R. L.~M.}
  \& \bibinfo{author}{Mampaso, A.}
\newblock \bibinfo{title}{Low-ionization structures in planetary nebulae:
  Confronting models with observations}.
\newblock \emph{\bibinfo{journal}{The Astrophysical Journal}}
  \textbf{\bibinfo{volume}{547}}, \bibinfo{pages}{302--310}
  (\bibinfo{year}{2001}).

\bibitem{richer08}
\bibinfo{author}{Richer, M.~G.}
\newblock \bibinfo{title}{Optical line profiles of the helix planetary nebula
  (ngc 7293) to large radii}.
\newblock \emph{\bibinfo{journal}{Monthly Notices of the Royal Astronomical
  Society}} \textbf{\bibinfo{volume}{384}}, \bibinfo{pages}{497--503}
  (\bibinfo{year}{2008}).

\bibitem{vandesteene15}
\bibinfo{author}{Van~de Steene, G.~C.}, \bibinfo{author}{van Hoof, P. A.~M.},
  \bibinfo{author}{Exter, K.~M.}, \bibinfo{author}{Barlow, M.~J.} \emph{et~al.}
\newblock \bibinfo{title}{Herschel imaging of the dust in the helix nebula (ngc
  7293)}.
\newblock \emph{\bibinfo{journal}{Astronomy \& Astrophysics}}
  \textbf{\bibinfo{volume}{574}}, \bibinfo{pages}{A134} (\bibinfo{year}{2015}).

\bibitem{maskfill}
\bibinfo{author}{{van Dokkum}, P.} \& \bibinfo{author}{{Pasha}, I.}
\newblock \bibinfo{title}{{A Robust and Simple Method for Filling in Masked
  Data in Astronomical Images}}.
\newblock \emph{\bibinfo{journal}{\pasp}} \textbf{\bibinfo{volume}{136}},
  \bibinfo{pages}{034503} (\bibinfo{year}{2024}).
\newblock \eprint{2312.03064}.

\bibitem{keim:22}
\bibinfo{author}{{Keim}, M.~A.} \emph{et~al.}
\newblock \bibinfo{title}{{Tidal Distortions in NGC1052-DF2 and NGC1052-DF4:
  Independent Evidence for a Lack of Dark Matter}}.
\newblock \emph{\bibinfo{journal}{\apj}} \textbf{\bibinfo{volume}{935}},
  \bibinfo{pages}{160} (\bibinfo{year}{2022}).
\newblock \eprint{2109.09778}.

\bibitem{Jin:25}
\bibinfo{author}{{Jin}, Y.} \& \bibinfo{author}{{Raymond}, J.}
\newblock \bibinfo{title}{{Dialog Concerning the Two Shock Codes}}.
\newblock \emph{\bibinfo{journal}{\apj}} \textbf{\bibinfo{volume}{989}},
  \bibinfo{pages}{203} (\bibinfo{year}{2025}).
\newblock \eprint{2507.03225}.

\bibitem{dfproject}
\bibinfo{author}{{Rhea}, C.} \emph{et~al.}
\newblock \bibinfo{title}{{dfreproject: A Python package for astronomical
  reprojection}}.
\newblock \emph{\bibinfo{journal}{The Journal of Open Source Software}}
  \textbf{\bibinfo{volume}{10}}, \bibinfo{pages}{8525} (\bibinfo{year}{2025}).
\newblock \eprint{2505.03932}.

\end{thebibliography}

\renewcommand\thefigure{Extended Data Figure \arabic{figure}}
\setcounter{figure}{0}

\begin{figure*}[ht]
  \begin{center}
  \includegraphics[width=0.9\linewidth]{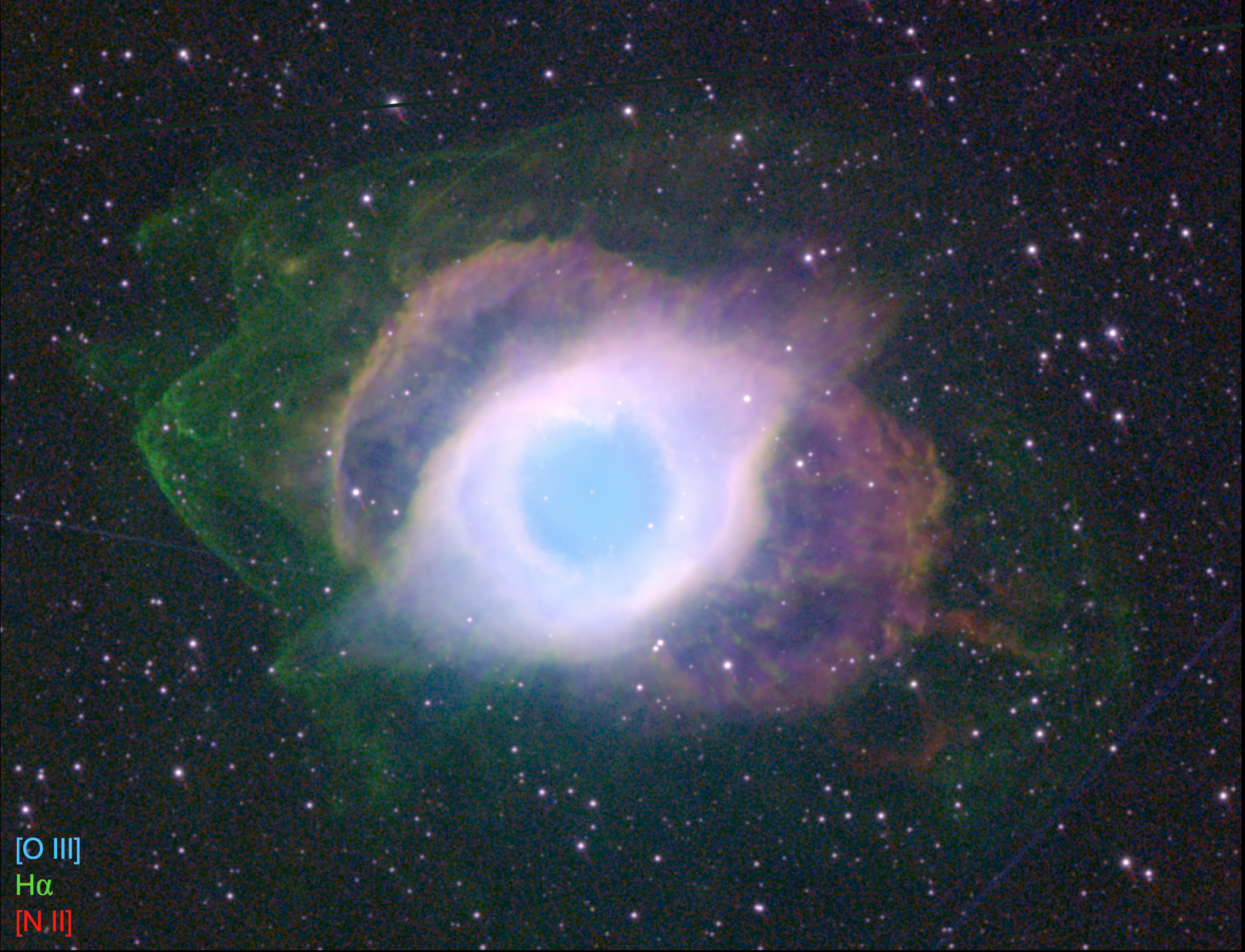}
  \end{center}
    \vspace{-0.3truecm}  
    \caption{\small \textbf{Color representation of the Helix Nebula.} 
Combination of MOTHRA [O\,III], H$\alpha$, and [N\,II] imaging of the
Helix. The bow shocks on the eastern side of the nebula, as well as the weak bows and bubbles
in the west, show up as
green in this representation: they are relatively
bright in H$\alpha$ and faint in [O\,III] and [N\,II].
   }
   \label{col.fig}
    \vspace{-12pt}
\end{figure*}

\begin{methods}

\subsection{Observations\vspace{0.3cm}\\}

The Helix Nebula was observed as a
commissioning target during the early construction phase of
MOTHRA. When completed, the array will comprise 1140 Canon 400\,mm f/2.8 II/III
telephoto lenses distributed over 28 narrow band mounts and 2
broad band mounts. Each narrow band mount has 38 lenses: 18
are equipped with an H$\alpha$ filter, 8 with [O\,III]\,$\lambda 5007$,
8 with [N\,II]\,$\lambda 6583$, and 4 are continuum filters. The widths of the ultra-narrow
interference filters are $0.71$\,nm for [O\,III] and $0.93$\,nm
for H$\alpha$ and [N\,II], corresponding to a velocity bandwidth of
$\approx 430$\,\kms.
The continuum filters, two for H$\alpha$ and [N\,II] and two for [O\,III],
are $\approx 90$\,nm wide with a $\approx 33$\,nm
notch in the middle at the location of the emission lines. These
filters therefore measure the continuum at the wavelengths of the lines, without
being contaminated by them.

The observations reported here were obtained on November 16 and November 19--24 2025,
when the first five mounts of MOTHRA were operational.
As the number of active lenses varied during the observations
we express the exposure time as single-lens
equivalent (SLE) hours, that is, the exposure time multiplied by the
number of lenses that were in use during the observation.
In H$\alpha$ the on-target
exposure time was 172.5 SLE hours, corresponding to 20 minutes with the
504 H$\alpha$ lenses that the completed MOTHRA will have.

\subsection{Data Reduction\vspace{0.3cm}\\}

The data reduction includes several steps that are unique to
the design of MOTHRA and its prototype, the Dragonfly Spectral Line
Mapper.\cite{lokhorst:24,chen:25}
The $\approx 400$\,\kms\ wide
interference filters are placed in front of the lenses,
and specific wavelengths are chosen
by tilting the filters. Owing to the large
($\approx 3^{\circ}$) field of view
the wavelength of the bandpass is not constant over the image
but varies as a function of the angular displacement of the target relative to the projected tilt axis of the filter, as well as the angular distance
to the optical axis.
As a result, the sky background is complex:
sky emission lines appear as broad bands in the image,
particularly at the large tilts that are used for Galactic objects.\cite{lokhorst:24}
For the Helix observations we obtained offset sky exposures in a circular
pattern around the science field to model the background. Bracketing
sky exposures were combined and subtracted from each individual science frame,
prior to combining the science data.

A color image of the Helix is shown in Fig.\ \ref{col.fig}, created from
the [O\,III], H$\alpha$, and [N\,II] data.  
It is well established from narrowband imaging and line-ratio mapping that the
morphology of PNe
varies strongly between different emission lines, with low-ionization structures often enhanced in [N\,II] relative to [O\,III].\cite{balick87,corradi96,goncalves01}
The Helix shows the same behavior in its central regions and faint outer halo.\cite{odell:04,richer08,vandesteene15}
In the MOTHRA images, the forest of bow shocks is relatively bright
in H$\alpha$ and is faint in [O\,III] and [N\,II].

The final step in the data reduction is the subtraction of continuum emission
from the narrow band data;
this was done by matching the point spread functions (PSFs) of the two images
and scaling the continuum image to match the normalization of the narrow band
image. Remaining residuals were filled in with the {\tt maskfill} code.\cite{maskfill}
The depth of the images was determined with the {\tt sbcontrast} method,
which empirically determines the contrast sensitivity on a desired spatial
scale.\cite{keim:22} The reduced and sky-subtracted narrow band images
reach a $1\sigma$ depth of 
$\approx 2\times 10^{-19}$\,erg\,s$^{-1}$\,cm$^{-2}$
in H$\alpha$ on $1'$ scales.

%

\subsection{Shock velocity\vspace{0.3cm}\\}

To interpret the observed emission-line ratios we compute a grid of radiative shock models using the MAPPINGS~V code.\cite{Sutherland2018} The models are run with Solar abundances and include a self-consistent treatment of the photoionizing precursor, in which the upstream ionization and temperature structure are iteratively determined from the radiation field produced in the post-shock cooling zone.
We adopt a pre-shock hydrogen density of $n_{\rm H} = 5~\mathrm{cm^{-3}}$, consistent with the fiducial density used in recent MAPPINGS~V benchmark calculations\cite{Jin:25}, and appropriate for diffuse circumstellar and interstellar environments. The results are insensitive to the choice of $n_{\rm H}$, as the lines form when much higher densities
are reached.
The magnetic field strength is parameterized through $\alpha \equiv B/\sqrt{n_{\rm H}}$ (with $B$ in $\mu$G and $n_{\rm H}$ in cm$^{-3}$), which controls the degree to which magnetic pressure limits compression in the post-shock cooling zone.\cite{Sutherland2018}

We explore two representative regimes: a weak-field case with $\alpha = 1$, which effectively captures the hydrodynamic limit (the results are virtually identical for all
values $\alpha \lesssim 1$), and a magnetically supported case with $\alpha = 2$, for which magnetic pressure significantly reduces the maximum compression. 
Shock velocities are sampled finely in the range $v_{\rm shock} =20-115$\,\kms\ to resolve the transition in ionization structure associated with the onset of [N\,II] and
[O\,III] emission.

\begin{figure}[ht]
  \begin{center}
  \includegraphics[width=0.90\linewidth]{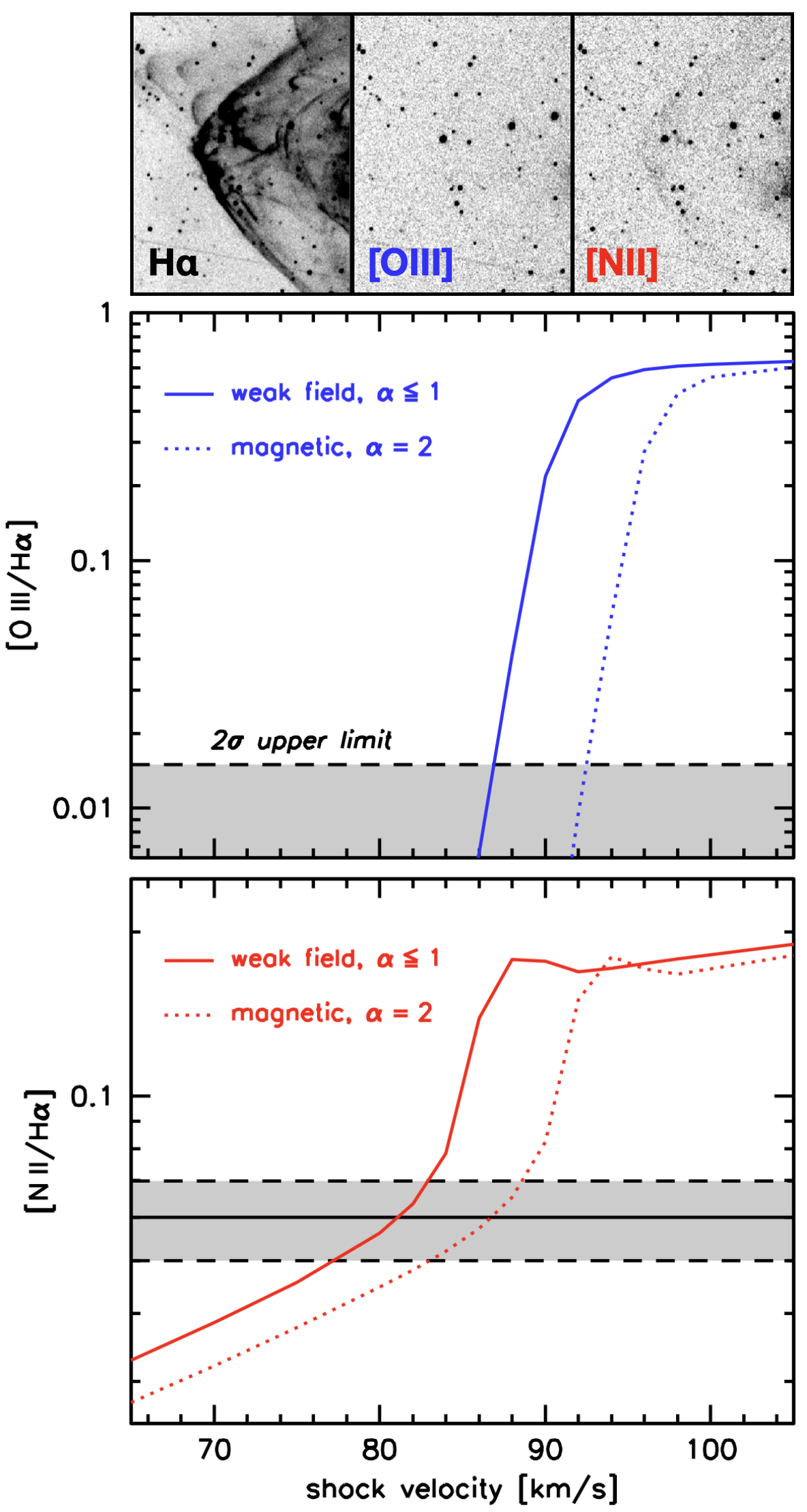}
  \end{center}
    \vspace{-0.5truecm}  
    \caption{\small \textbf{Shock velocities from line ratios.} 
The measured [O\,III] and [N\,II] line ratios for the brightest bow (top),
compared to MAPPINGS~V models of shocks propagating in a largely neutral medium.
Solid lines show the weak field limit with
$\alpha\lesssim 1$ and broken lines show a model with a moderately strong
magnetic field. The [N\,II] line strength increases gradually with shock velocity,
whereas [O\,III] turns on rapidly above a threshold.
The observed line ratios indicate shock velocities of $80-90$\,\kms,
depending on the magnetic field strength.
   }
   \label{line_ratios.fig}
    \vspace{-0pt}
\end{figure}

\begin{figure*}[ht]
  \begin{center}
  \includegraphics[width=1.0\linewidth]{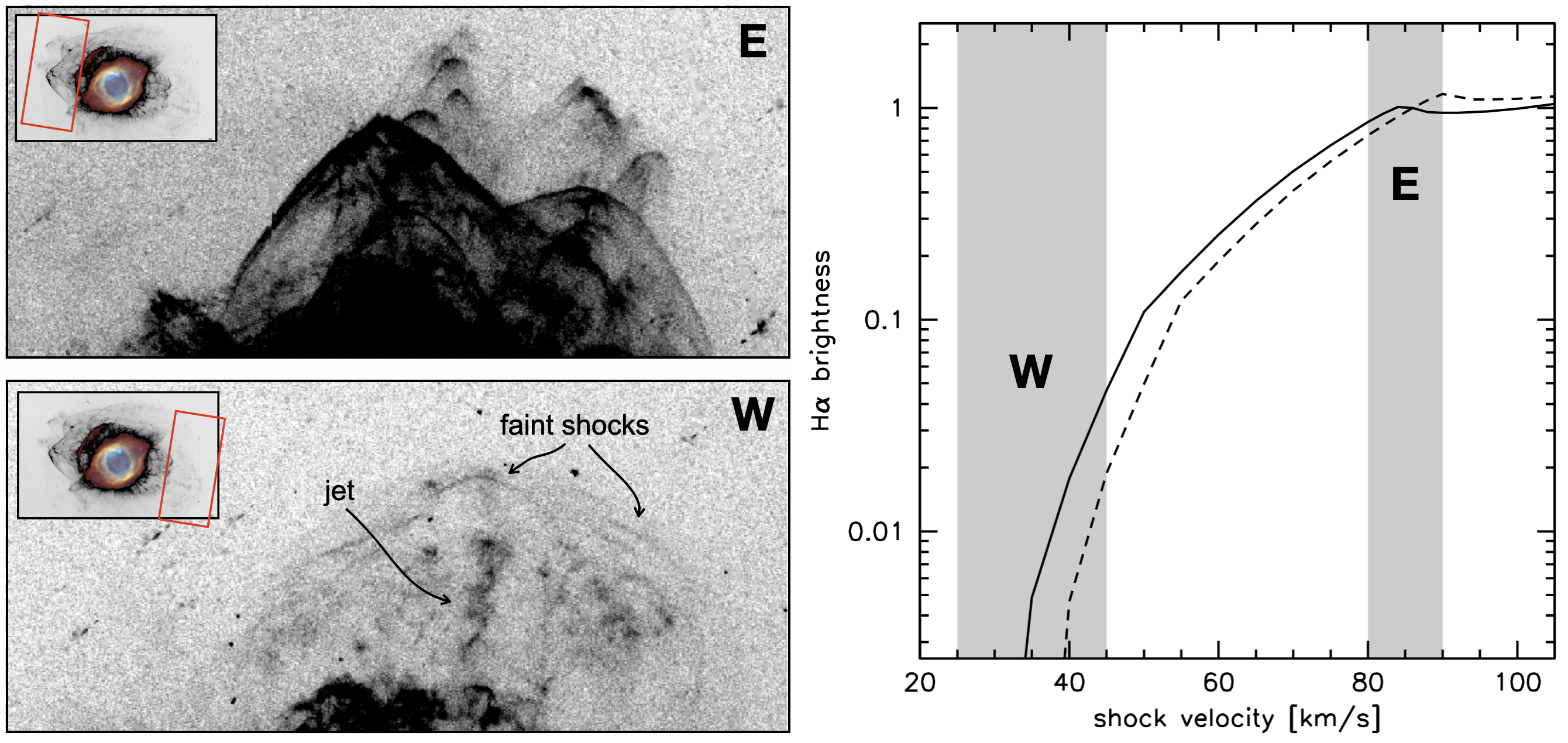}
  \end{center}
    \vspace{-0.5truecm}  
    \caption{\small \textbf{Shocks on the East and West side.} 
Both the east and west side of the Helix show extended regions of H$\alpha$ emission,
extending beyond where [O\,III] and [N\,II] are detected. {\em Left panel:}
Side by side comparison of the two opposing regions,
along an axis of 80$^{\circ}$ measured N through E.
There are faint bubbles and shocks in the west, that we interpret as possible counterparts
of the shocks in the east. {\em Right panel:} H$\alpha$ luminosity as a function
of shock velocity, normalized at $v_{\rm shock}=85$\,\kms,
in the MAPPINGS~V code.\cite{Sutherland2018} The solid line is for weak
magnetic fields ($\alpha \leq 1$) and the broken line for moderately strong magnetic
pressure ($\alpha = 2$). Grey bands indicate the inferred shock velocities on
each side of the nebula: $85 \pm 5$\,\kms\ on the east side
and $35 \pm 10$\,\kms\ on the west side (where the ambient
flow is assumed to be $5\pm 5$\,\kms). The H$\alpha$ luminosity is expected to be 1--2
orders of magnitude fainter in the west than in the east, based on shock velocity alone.
   }
   \label{east_west.fig}
    \vspace{-12pt}
\end{figure*}

The results are shown in Fig.\ \ref{line_ratios.fig}. The line ratios are a strong function
of shock velocity in this regime: the post-shock temperature increases as
$T \propto v_{\rm shock}^2$, leading to a rapidly increasing flux of ionizing photons
from the cooling zone and its associated precursor.\cite{Sutherland2018}
For [O\,III],
the onset is abrupt once the radiation field becomes sufficiently hard to sustain
an extended $\mathrm{O}^{++}$ zone, whereas lower-ionization species such
as [N\,II] respond more gradually.
The observed upper limit on [O\,III]/H$\alpha$ and measured [N\,II]/H$\alpha$ (the grey
bands in Fig.\ \ref{line_ratios.fig}) indicate
a shock velocity of $v_{\rm shock} =80-90$\,\kms\ depending
on the magnetic field strength, and we use $v_{\rm shock}=85\pm 5$\,\kms\ in the main
text.

\begin{figure*}[ht]
  \begin{center}
  \includegraphics[width=0.9\linewidth]{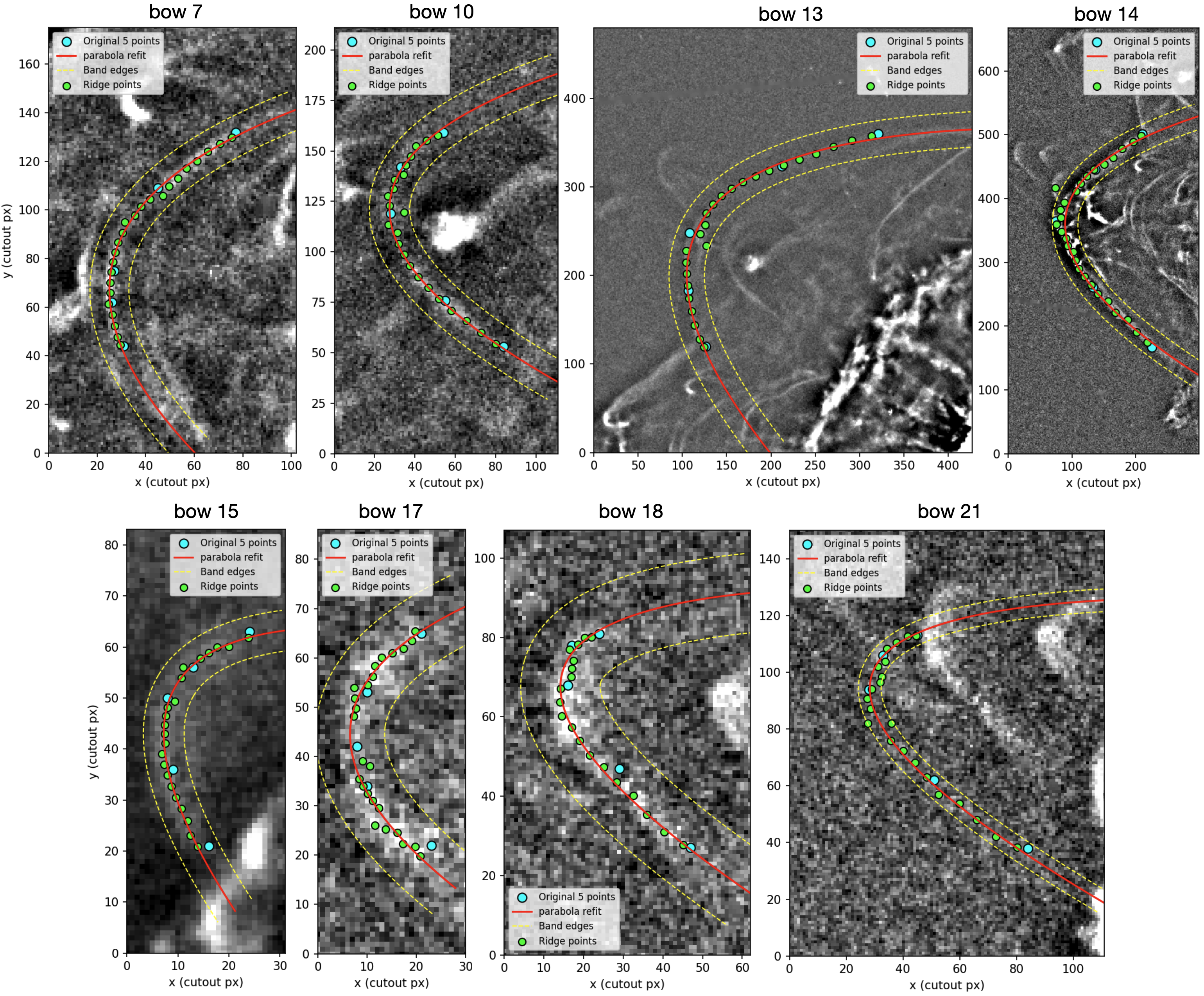}
  \end{center}
    \vspace{-0.3truecm}  
    \caption{\small \textbf{Examples of profile fits.} 
Eight bows covering a range of distances from the white dwarf are shown, in
unsharp masked H$\alpha$ images.
For each bow, blue points show the initial manual definition of the structure.
A polynomial fit to the blue points is used to determine the initial fitting region.
Green points show measurements of the ridge line, that is, the peak of emission within
the yellow fitting region. Red curves are parabolic profile fits to the green points,
not taking the blue points into account. The pixel size is $2''$.
   }
   \label{fit_examples.fig}
    \vspace{-12pt}
\end{figure*}

\subsection{H$\alpha$ luminosity as a function of shock velocity\vspace{0.3cm}\\}

There is a clear asymmetry in the brightness of shocks in the outer Helix nebula,
with all strong shocks located on the eastern side. As can be seen in
Figs.\ \ref{helix_ha_large.fig} and \ref{col.fig} There does appear to be a corresponding
region on the other side of the nebula, showing rounded arcs and bubbles
that are only detected in H$\alpha$. In Fig.\ \ref{east_west.fig} we show a side by side
comparison of the two regions.
Here we ask whether the difference in brightness
between the eastern and western shocks can be explained by the difference of the
shock velocities.

The right panel of Fig.\ \ref{east_west.fig} shows the H$\alpha$ luminosity as a
function of shock velocity, from the MAPPINGS~V code.\cite{Sutherland2018} 
At low velocities, the H$\alpha$ luminosity rises steeply with shock velocity as the shock progressively ionizes neutral hydrogen. At higher velocities, once the radiative precursor pre-ionizes the upstream gas, the hydrogen ionization fraction approaches unity and the
H$\alpha$ emissivity no longer tracks the mechanical energy flux. Instead, additional energy is channeled into heating and metal-line cooling, resulting in a flattening of the
H$\alpha$–velocity relation.

Grey bands show the inferred shock velocities, $85\pm 5$\,\kms\
on the east side and $35\pm 10$\,\kms\ on the west side (see main text).
Because of the steep rise of the H$\alpha$\,--\,$v_{\rm shock}$ relation
in this regime, the H$\alpha$
luminosity is expected to be 1--2 orders of magnitude higher on the east side than
on the west side. The difference in the
observed surface brightness is a factor of $\sim 10$, broadly consistent with the
shock models.

We note that the H$\alpha$ features on the west side are probably not
related to the
high velocity jet in that region, even though they appear to be in front
of it in projection (see Fig.\ \ref{east_west.fig}). The jet has
a velocity of $\sim 300$\,\kms, measured directly from H$\alpha$ and
[N\,II] spectroscopy,\cite{meaburn:13} and
if the shocks were driven by the jet they would show strong [O\,III] emission.

\subsection{Fitting of bow shocks\vspace{0.3cm}\\}

The bows are fit in the following way.
A bow is initially characterized by
five manually-selected ($x$,$y$) positions:
two marking the end points of the visible arc,
one marking the approximate apex, and two points in between the end points and
the apex. Next, a second-order polynomial is fit to these five points. This
fit provides a reasonable approximation of the area of the image where the bow
is located. A band of typical width $\pm 20$ pixels is defined around the
polynomial fit. Within this band the
ridge of the bow is identified by determining the local maximum in bins along
the polynomial. To enhance contrast,
an unsharp-masked version of the H$\alpha$ image is used to measure the ridge.

With the ridge line measurements in place, 
a fit to an analytic function is
performed. Following earlier work,\cite{tarango:18} we fit four families of
functions: parabolas, hyperbolas, ellipses, and Wilkinoids.\cite{Wilkin1996}
For each functional form we determine the associated radius of curvature $R_{\rm c}$
and the related distance between the apex and the object $R_0$. 
For a parabola $R_0 = R_{\rm c}/2$; for a hyperbola and an ellipse $R_0 =
R_{\rm c}/(1+e)$, with $e$ the ellipticity, and for a Wilkin profile
$R_0 = 3 R_{\rm c}/5$.
The five manually selected points are not used
in the fit. After the initial fit the band is redefined, now based
on the specific analytic function rather than the polynomial, and the ridge line measurements
and fit are repeated.
The process is illustrated for eight of the bows in Fig.\ \ref{fit_examples.fig}.
The fits are stable for the 22 bows that we identify in the image.

As discussed in the main text, we use the parabola as the fiducial profile, and
take the rms ranges in $R_{\rm c}$ and $R_0$ from the four fits as uncertainties.

\subsection{Data availability.}
The reduced, continuum-subtracted H$\alpha$ image is available at Zenodo
(https://doi.org/10.5281/zenodo.19864496).

\subsection{Code availability.}
We have made use of standard data analysis tools in the Python
environment. Image projections were performed with {\tt dfproject},\cite{dfproject}
available at https://github.com/DragonflyTelescope/dfreproject.

\end{methods}

\vspace{3pt}
\noindent\rule{\linewidth}{0.4pt}
\vspace{3pt}

\begin{addendum}
 \item[Acknowledgements]
We thank the staff at Obstech observatory for their invaluable help with the
construction and operation of MOTHRA. The quality of the
manuscript benefitted greatly from the insightful comments of William Henney and
the other, anonymous, reviewer(s).

\item[Funding Statement]
MOTHRA is made possible by the funding and ongoing support from Alex Gerko,
Founder and CEO of XTX Markets.

 \item[Author Contributions] P.v.D., R.A., W.P.B., S.C., S.R.J., D.M.L., I.P.\ and
 C.R.\ contributed to the
construction of the array, the execution of the observations and the data reduction. P.v.D.\
led the analysis and wrote the paper. R.A.\ aided in the interpretation.

\item[Competing Interests]
The authors declare that they have no
   competing financial interests. 
 \item[Additional Information] Correspondence and requests for
   materials should be addressed to P.v.D.~(email:
   pieter.vandokkum@dragonfly1000.com).

\end{addendum}

\appendix

\end{document}